\documentclass[12pt]{article}
\usepackage{amssymb}
\usepackage{a4}
\evensidemargin \oddsidemargin
\def\1{{\bf 1}}
\def\id{\mbox{id\,}}
\def\ot{\!\otimes\!}

\def\F{\mbox{$\cal F$}}
\def\Fu{{\cal F}^{\alpha}}
\def\Fd{{\cal F}\!_{\alpha}}
\def\bFu{\overline{{\cal F}}^{\alpha}}
\def\bFd{\overline{{\cal F}}\!\!_{\alpha}}
\def\bF{\mbox{$\overline{\cal F}$}}

\def\ra{\rangle}
\def\la{\langle}
\def \A{{\cal A}}
\def \B {{\cal B}}

\def \D {{\cal D}}

\def \H {{\cal H}}
\def \bH {\overline{\cal H}}

\def \O {{\cal O}}
\def \P {{\cal P}}

\def \X {{\cal X} }
\def \rx{{\rm x}}
\def \hrx{\hat{\rm x}}

\def \hau{{\sf h}}
\def \haus{{\sf h}_{\scriptscriptstyle \star}}
\def \hhau{\hat {\sf h}}
\def \Ha{{\sf H}}

\def \Han{{\sf H}^{\scriptscriptstyle (n)}}
\def \Hans{{\sf H}^{\scriptscriptstyle (n)}_{\scriptscriptstyle \star}}
\def \Has{{\sf H}_{\scriptscriptstyle \star}}

\def \psin{\psi^{\scriptscriptstyle (n)}}
\def\R{\mbox{$\cal R$}}
\def\Ru{{\cal R}^{\alpha}}
\def\Rd{{\cal R}_{\alpha}}

\def\hA{\mbox{$\widehat{\cal A}$}}

\def\hH{\mbox{$\hat{H}$}}

\newcommand{\trc}{\triangleright}
\def\g{\mbox{\bf g\,}}

\def\nn{\nonumber \\}

\newcommand{\be}{\begin{equation}}
\newcommand{\ee}{\end{equation}}
\newcommand{\bea}{\begin{eqnarray}}
\newcommand{\eea}{\end{eqnarray}}
\newcommand{\ba}{\begin{array}}
\newcommand{\ea}{\end{array}}

\begin{document}
\title{Learning from Julius' star, $*$, $\star$}

\author{   Gaetano Fiore\footnote{Talk given at the SEENET-MTP Workshop
``Scientific and Human Legacy of Julius Wess'' (JW2011), 
Donji Milanovac (Serbia), 27-28 August 2011, within the
Balkan Summer Institute 2011.  To appear in the proceedings.
},  \\\\
Dip. di Matematica e Applicazioni, Universit\`a ``Federico II''\\
   V. Claudio 21, 80125 Napoli, Italy\\         
I.N.F.N., Sez. di Napoli,
        Complesso MSA, V. Cintia, 80126 Napoli, Italy}
\date{}

\maketitle

\begin{abstract}
\noindent
While collecting some personal memories about Julius Wess,
I briefly describe some aspects of my recent work on 
many particle quantum mechanics and second quantization on noncommutative 
spaces obtained by twisting, and their connection to him.  

\end{abstract}

\bigskip
\noindent
{\it Keywords: Noncommutative spaces; Drinfel'd twist; second quantization. }

\section{Introduction}

I'll try to sketch Julius' direct or indirect impact on my life and scientific activity.
I have learnt a lot  from his work (I wish I could have learnt more!   Reading his papers I still learn),
especially on some issues  he was never tired of emphasising, such as the 
importance of symmetries (groups, supergroups, quantum groups,...)  
and conservation laws in physics, but also his guiding idea 
that fundamental physical laws should be coincisely expressible in
algebraic form (sometimes he joked: ``At the Very Beginning There Was the Algebra...''). 
I have learnt also from him through
his scientific and human qualities (they often overlapped).
Among them I would certainly mention:
physical intuition and ``exploration sixth sense'';  open-minded, 
independent and creative thinking; search for beauty and simplicity;
hierarchy of arguments, conciseness, clarity; 
concreteness, honesty, humbleness; ambition, courage to dare; 
coherence, rigour, determination;
familiarity, cosiness; kindness, elegance, sense of humour.

I first learned about Prof. J. Wess at Naples University during my degree thesis
on BRST quantization of gauge theories, when I met the {\it Wess-Zumino consistency 
equation} for the anomaly. After the degree I read with enthusiasm his  seminal 
paper Ref. \cite{WesZum90} with Zumino,
which marked the beginning of their work  on quantum spaces and 
quantum groups.
Spured by my advisor M. Abud, in early 1990 I sent him a letter asking whether 
I could do my PhD under his guide. As customary, he answered he would accepted me, but 
could not provide financial support. Later that year I won a PhD grant at SISSA
and  started my PhD there. It was a very stimulating environment, nevertheless
I kept the eye on what Julius and his group were doing, and in 1991 my Master 
and PhD supervisor L. Bonora accepted that I would do my theses on the
same topics. Soon I succeded in finding a sensible definition of integration over
the socalled quantum Euclidean space $\mathbb{R}_q^n$ and in solving the eigenvalue problem
for the harmonic oscillator Hamiltonian on $\mathbb{R}_q^n$, which Wess had proposed 
to me; later I succeded in realizing $U_qso(n)$ (the deformed infinitesimal rotations)
by differential operators  on $\mathbb{R}_q^n$ (the analog of the angular momentum
components)\cite{Fio95cmp}. In summer 1993 I wished to update him about the progresses,
but he was very busy and difficult to meet. I remember that as a last year Sissa student I could 
participate to one conference outside Europe, and felt a strong appeal  towards the
``First Caribbean School of Mathematics and Theoretical 
Physics Saint-Fran\c{c}ois, Guadeloupe''. But I decided to go to the Workshop 
``Interface between physics and  mathematics'' in Hangzhou, China, after 
noticing Julius among the invited speakers. It was an interesting conference 
and a marvellous trip, but Julius was at some other conference. I thus learned what being a 
scientific ``Star'' like Julius meant: I soon realized that the 
\medskip {}
 \noindent
\begin{tabular}{l l} \hskip-.2cm
\parbox[l]{7cm}{
www.ysfine.com confmenu website was full of announcements of overlapping conferences
like the ones beside: the symbol ``*'' had become a sort of royal 
crown over his name. Was he {\bf too kind} to say
a clear ``No'' to the insistent invitation of conference organizers, or {\bf so}
{\bf open} that he did not  exclude accepting the  invitation at 
the last moment?} \ \ \ \begin{tabular}{|l|l|}
\hline
Blabla Conference & Blabla Symposyum \\[8pt] 
\hline
9-13/9/yyyy & 10-15/9/yyyy \\ 
Speakers include: & Speakers include: \\
-  Julius Wess* & -  Julius Wess* \\
-  ... & -  ... \\
-  ... & -  ...  \\
-  ... & -  ...  \\
\hline
* To be confirmed & * To be confirmed \\\hline
\end{tabular}
\end{tabular}
\vskip.1cm 
\noindent
I still don't know the right answer.
After my PhD I was hosted by Julius as a A. v. Humboldt post-doc at the
{\it Ludwig-Maximilian-Universit\"at} in Munich. At the time the symbol ``*''
recurred obsessively in our computations for a different reason: we were struggling
with $*$-structures (i.e. the algebraic formulations of hermitean 
conjugations). Julius and his group we could not find (for real 
deformation parameters $q$) $*$-structures compatible with the
$q$-Poincar\'e (nor the $q$-Euclidean) quantum group without doubling the
generators of translations and adding dilatations. I appreciated first his rigorous
and hard working, finally his intellectual honesty in admitting that fact made
those deformations unsatisfactory and to quit. 
Ever since I have worried about implementing $*$-structures in the
noncommutative world.

In Munich I also learnt  (especially thanks to his student R. Engeldinger)
about the construction of quantum groups from groups using
{\it Drinfel'd twists} $\F$\cite{Dri}; 
among other things $\F$ intertwined between the 
group and the quantum group actions on tensor product representations. 
In the joint papers Ref. \cite{FioSch96} with Peter Schupp we pointed out that 
the unitary transformation $\F$ (and its descendants) intertwines also between 
the conventional and an unconventional realization of the permutation group 
[and therefore of (anti)symmetrization] 
on tensor products, and therefore that quantum group transformations were compatible
with Bose and Fermi statistics. 
Here I would like to sketch some related, more recent results\cite{Fio10} 
 illustrating the crucial role of twists not only in deforming spaces but also 
in quantizating (for simplicity scalar) fields on the latter.
This will clarify also the last symbol $\star$ in the title. 

 A rather general way to
deform an (associative) algebra $\A$ (over $\mathbb{C}$, say) into a new one $\A_\star$
is by deformation quantization\cite{BayFlaFroLicSte}. Calling
$\lambda$ the deformation parameter, this means that the two have
the same vector space over $\mathbb{C}[[\lambda]]$,
$V(\A_\star)=V(\A)[[\lambda]]$, but the product $\star$ in
$\A_\star$ is a deformation of the product $\cdot$ in $\A$. On the
algebra $\X$ of smooth functions on a manifold $X$, and on the
algebra $\D\supset\X$ of  differential operators on $\X$,
$f\star h$  can be defined applying to $f\ot h$ first a suitable
bi-pseudodifferential operator $\bF$ (depending on the deformation
parameter $\lambda$ and reducing to the identity when $\lambda=0$)
and then the pointwise multiplication $\cdot$. The simplest example
is probably  the Gr\"onewold-Moyal-Weyl
(Moyal, for brevity) $\star$-product on $X=\mathbb{R}^m$:
\be 
\ba{l} a(x)\!\star\! b(x)\!:=\! a(x) \exp\left[\frac
i2\stackrel{\longleftarrow}{\partial_h}\!\lambda\vartheta^{hk}\!
\stackrel{\longrightarrow}{\partial_k}\right] b(x)=
\cdot\left[\bF(\trc\ot\trc)(a\ot b)\right],
\\[8pt] \bF
:= \mbox{exp}\left(-\frac i2\theta^{hk}P_{h}\ot
P_{k}\right),\qquad\qquad \theta^{hk}:=\lambda\vartheta^{hk},
\label{Moyalstarprod}\ea 
\ee 
where $P_a$ are the generators of translations (on $\X$
$P_a$ can be identified with $-i\partial_a:=-i\partial/\partial x^a$), and
 $\vartheta^{hk}$ is a fixed real antisymmetric matrix; as recalled below, definition (\ref{Moyalstarprod})$_1$ can be made non-formal in terms of  Fourier transforms.
$\X_\star,\X$ have the Poincar\'e-Birkhoff-Witt (PBW) property, i.e. the subspaces of
$\star$-polynomials and $\cdot$-polynomials of any fixed degree in
$x^h$  coincide. One can define a linear map
$\wedge:f\!\in\!\X[[\lambda]]\!\to\!\hat f\!\in\!\X_\star$  (the {\it Weyl map}) by requiring 
that it reduces to the identity on the vector space
$V(\X_\star)=V(\X)[[\lambda]]$: $\hat f({\rm x}\star)=f({\rm x})$. One finds
 \be \ba{l}
 \wedge(x^h)=x^h,\\
\wedge(x^hx^k)=x^h\star x^k-\frac i2\theta^{hk},
\qquad\Rightarrow\qquad [x^{h}\stackrel{\star}, x^{k}]=\1 i\theta^{hk},\quad\\
\qquad\qquad ...\label{defhatx} \ea 
\ee
and so on (again, this can be extended to non-polynomial functions
through Fourier transforms). In other words, by $\wedge$ one expresses 
functions of ${\rm x}^h$ as
functions of ${\rm x}^h\star$. As $\X$, also $\X_\star$ can be defined purely through generators
and relations: the coordinates ${\rm x}^h$ and $\1$ are the generators of both, 
and fulfill $[x^{h}, x^{k}]\!=\!0$ in $\X$, (\ref{defhatx})$_2$ in $\X_\star$. 
Similarly one deforms  $\D$ into $\D_\star$; however for the twist (\ref{Moyalstarprod})
$[\partial_a\stackrel{\star},  \cdot]=[\partial_a, \cdot]$.

\noindent
Replacing all $\cdot$ by $\star$'s e.g. in  the Schr\"odinger equation 
of a particle with charge $q$
\be
\ba{c}  \haus\psi({\rm x})=i\hbar  \partial_t
\psi({\rm x}),\quad \qquad
\haus:=\big[\!\frac{-\hbar^2}{2m}D_a\!\star\!
D_a\!+\!V\big]\star,\quad \qquad D_a\!=\!\partial_a\!+\! i \frac{q}{\hbar c}A_a,
\ea                  \label{1Schr}
\ee
we obtain a pseudodifferential equation and therefore introduce some 
(quite special) non-local interaction, which might e.g. give an effective 
description of a complicated background.
The use of noncommutative coordinates may then help to solve the dynamics:
if we express $V(\rx)\star, \: A_a(\rx)\star$ and $\psi({\rm x})$ as their Weyl map images
$\hat V(\rx\star),\: \hat A_a(\rx\star)$ and $\hat\psi({\rm x}\star)$, then  (\ref{1Schr}) becomes 
a {\it second order $\star$-differential} equation
(i.e. of second degree in $\partial_h\star$) where the unknown is
now a funtion $\hat\psi({\rm x}\star)$ of ${\rm x}\star$:
$$
\ba{l}
i\hbar  \partial_t \hat\psi(\hrx)=
\frac{-\hbar^2}{2m}\hat D_a\hat D_a\hat\psi(\hat{\rm
x})\!+\! \hat V(\hat{\rm x})\hat\psi(\hat{\rm x})
;\ea
$$
here we have made the notation lighter by denoting \ $\rx^a\star,\:\partial_a\star$ as
$\hrx^a,\: \hat\partial_a$. (More generally, we often change notation as follows:
 \ \ $\X_\star\leadsto\hat \X$,
$\D_\star\leadsto\hat \D$, $x^h_j\!\star\! \leadsto \hat x^h_j$,
$\partial_h^j\!\star\!\leadsto \hat \partial_h^j$, $a^+_i\!\star\!
\leadsto \hat a^+_i$, etc.).
Nonetheless, 
in a conservative approach we still measure the position of a particle using
the observables $\rx^a$ of {\it commutative} space. In a radical one
we will rather use the noncommutative observables $\hrx^a$
for the latter purpose.

Eq. (\ref{Moyalstarprod}-\ref{defhatx}) on Minkowski (resp. Euclidean) space are not 
covariant under the Poincar\'e (resp. Euclidean) group $G^e$,
or equivalently under the associated universal enveloping algebra $U\g^e$.
Julius and coworkers in Ref. \cite{Wes04,KocTso04}, simultaneously to Ref. \cite{ChaKulNisTur04},  
realized that $\F\!:=\!\bF^{-1}$ could be used to twist  $U\g^e$ into the 
symmetry Hopf  algebra of $\widehat{U\g^e}$ of  the $\star$-product itself: one could recover 
Poincar\'e covariance in a deformed form! In the joint 
paper  \cite{FioWes07} with Julius we pointed out that
in fact  $\widehat{U\g^e}$-covariance implied as 
$\star$-commutation relations for $n$ copies of $\X_\star$
 \be
[ x^{\mu}_i\stackrel{\star},x^{\nu}_j]=i\theta^{\mu\nu}
\quad\qquad\Leftrightarrow\quad\qquad  [\hat x^{\mu}_i,\hat x^{\nu}_j]=i\theta^{\mu\nu} \qquad\quad i\!=\!1,\!2,\!...,n,
\label{summary}
\ee
and not the ones with $i\delta_i^j\theta^{\mu\nu}$ at the rhs:
so for $i\!\neq \!j$ the rhs is not automatically zero. That has important consequences
for  both multiparticle quantum mechanics [$x_i$ denoting
the space(time) coordinates of the $i$-th particle]  and quantum field theory
(products of fields evaluated at $n$ different  spacetime 
points $x_i$). In Ref. \cite{FioWes07,Fio08Proc}  we found the surprising result that
{\it a translation-invariant Lagrangian implied that the field commutators
and Green functions, as functions of the coordinates' differences,  
remained as those of the undeformed theory} (at least for scalar fields).
To put the field quantization prescription on a firmer ground, in
Ref. \cite{Fio10} I have rederived it by a second quantization procedure from $n$-particle 
 wavefunctions preserving  Bose/Fermi statistics (i.e. the rule to compute the number 
of allowed states of $n$ identical bosons/fermions):
not only the function algebra, but also that of creation and annihilation 
operators and their tensor product are $\star$-deformed. 
[Following Julius, to guess the deformed analog of a known theory (be it noncommutative 
 gravity\cite{AscBloDimMeySchWes05,AscDimMeyWes06}
or gauge field theory\cite{AscDimMeyWes06LMP} before quantization, or  QFT) we should
translate all commutative notions into  their noncommutative analogs by just
expressing all products $\cdot$'s in terms of
$\star$-products.] 
I partly recall this in section \ref{se:2}, sticking to
the non-relativistic wave-mechanical formulation of  system 
of bosons/fermions on $\mathbb{R}^m$ and its second quantization;
as a new result I  point out that {\it if $\hau$ is not $G^e$-invariant, then 
the dynamics is deformed not only because $\haus\!\neq\! \hau$, but 
also because the $n$-particle Hamiltonian, beside $\sum_{i=1}^n\!\haus\!(\rx_i^a,\partial_{\rx_i^a})$, contains cross-terms}, so that the total energy is not additive, 
even if there are no explicit 2-particle interaction
terms. As preliminaries,
in section \ref{se:1} I briefly describe the twist-induced deformation of a cocommutative 
Hopf $*$-algebra and of its module $*$-algebras, in particular
 of the Heisenberg/Clifford algebra associated to 
bosons/fermions on $\mathbb{R}^m$ and  of the algebras
of functions and differential operators on  $\mathbb{R}^m$.

 \section{Preliminaries}\label{se:1}

\noindent
{\bf 2.1 \ Twisting $H\!=\!U\g$ to a noncocommutative Hopf
algebra $\hat H$.} 

The Universal Enveloping  $*$-Algebra (UEA) $H\!:=\!U\g$ of the 
Lie algebra $\g$ of any Lie group $G$
is a Hopf $*$-algebra. We briefly recall first what this means. Let
$$
\ba{lll}
\varepsilon(\1)=1,\qquad \quad &\Delta(\1)=\1\ot\1,\qquad \quad & S(\1)=\1,\\[6pt]
\varepsilon(g)=0,\qquad \quad & \Delta(g)=g\ot\1+\1\ot g,\qquad
\quad &  S(g)=-g,\qquad \qquad \mbox{if }g\in\g; \ea
$$
$\varepsilon,\Delta$ are extended to all of $H$ as
$*$-algebra maps, $S$ as  a $*$-antialgebra map: \be \ba{lll}
\varepsilon:H\to\mathbb{C},\quad\qquad  & \varepsilon(ab)=\varepsilon(a)\varepsilon(b),\quad\qquad  & \varepsilon(a^*)=[\varepsilon(a)]^*,\\[8pt]
\Delta:H\to H\ot H,\quad \qquad  & \Delta(ab)=\Delta(a)\Delta(b),
\qquad
 & \Delta(a^*)=[\Delta(a)]^{*\ot *},\\[8pt]
S:H\to H,\quad\qquad  & S(ab)=S(b)S(a),\quad\qquad  &
S\left\{\left[S(a^*)\right]^*\right\}=a. \ea \label{deltaprop} \ee
The extensions of $\varepsilon,\Delta,S$ are unambiguous, as
$\varepsilon(g)=0$,
$\Delta\big([g,g']\big)=\big[\Delta(g),\Delta(g')\big]$,
$S\big([g,g']\big)=\big[S(g'),S(g)\big]$ if $g,g'\in\g$. The maps
$\varepsilon,\Delta,S$ are the abstract operations by which one
constructs the trivial representation, the tensor product of any two
representations and the contragredient of any representation,
respectively; $H$ equipped with  $*,\varepsilon,\Delta,S$
is a {\it Hopf $*$-algebra}. 
One can deform this Hopf algebra using  a {\it twist}\cite{Dri} (see also
\cite{Tak90}), i.e. an element $\F\!\in\! (H\ot
H)[[\lambda]]$ fulfilling \bea &&\F=\1\ot\1+ O(\lambda),\qquad
\qquad (\epsilon\ot\id)\F=
(\id\ot\epsilon)\F=\1,                 \label{twistcond}\\[6pt]
&&(\F\ot\1)[(\Delta\ot\id)(\F)]=(\1\ot\F)[(\id\ot\Delta)(\F)]=:\F^3.
                                           \label{cocycle}
\eea
Let  $H\!_s\!\subseteq\!H$  be the smallest Hopf
$*$-subalgebra such that  $\F\!\in\! (H\!_s\ot H\!_s)[[\lambda]]$,
\be
 \Fu  \ot\Fd \!:=\F,\qquad
\quad  \bFu  \ot\bFd  \!:=\F^{-1},\qquad
\quad\beta:=
\Fu  S\!\left(\Fd \right)\in H\!_s[[\lambda]]   \label{defbeta}
\ee
(sum over $\alpha$) be the tensor decompositions of $\F,\F^{-1}$, and $\hH\!=\!H[[\lambda]]$. 
We  assume $\lambda$  real and $\F$ {\it unitary} ($\F^{*\ot *}=\F^{-1}$),
implying $\beta^*=S(\beta^{-1})$. Extending the
product, $*,\Delta,\varepsilon,S$ linearly to the formal power
series in $\lambda$ and setting \be \hat\Delta(g) :=\F \Delta(g)
\F^{-1}, \qquad  \hat S(g):=\beta \, S(g) \,\beta^{-1} ,\qquad
\R:=\tau(\F)\F^{-1},\label{inter-2}
 \ee
one finds that the analogs of conditions (\ref{deltaprop}) are
satisfied and therefore $(\hH,*,\hat\Delta , \varepsilon,\hat S)$ is
a Hopf $*$-algebra deformation of the initial one. 
While $H$ is cocommutative, i.e.
$\tau\!\circ\!\Delta(g)\!=\!\Delta(g)$ where $\tau$ is the
flip operator [$\tau(a\ot b)\!=\!b\ot a$], $\hH$ is
triangular noncocommutative, i.e.
$\tau\!\circ\!\hat\Delta(g)\!=\!\R\hat\Delta(g)\R^{-1}$,
with unitary triangular structure $\R$ (i.e. 
$\R^{-1}\!=\!\R_{\scriptscriptstyle 21}\!=\!\R^{*\ot *}$). Correspondingly,
$\hat\Delta , \hat S$ replace $\Delta, S$ in the construction of the tensor product of any two
representations and the contragredient of any representation,
respectively.
Drinfel'd has shown\cite{Dri} that any triangular deformation of
the Hopf algebra $H$ can be obtained in this way (up to
isomorphisms).

Eq. (\ref{cocycle}), (\ref{inter-2}) imply the
generalized intertwining relation
$\hat\Delta^{(n)}(g)\!=\!\F^n\Delta^{(n)}(g)(\F^n)^{-1}$
 for the iterated coproduct. By definition
$$
\hat\Delta^{(n)}: \hH\to \hH^{\ot n},\qquad\Delta^{(n)}: H[[\lambda]]\to
(H)^{\ot n}[[\lambda]],\qquad\F^n\in (H\!_s)^{\ot n}[[\lambda]]
$$
reduce to $\hat\Delta,\Delta,\F$ for $n=2$, whereas for $n>2$ they
can be defined recursively as \be \ba{l}
\hat\Delta^{(n\!+\!1)}=(\id^{\ot(n\!-\!1)}\ot\hat\Delta)
\circ\hat\Delta^{(n)},\qquad\Delta^{(n\!+\!1)}=
(\id^{\ot{(n\!-\!1)}}\ot\Delta)\circ\Delta^{(n)},\\[8pt]
\F^{n\!+\!1}=(\1^{\ot{(n\!-\!1)}}\ot\F)[(\id^{\ot{(n\!-\!1)}}
\ot\Delta)\F^n].                \label{iter-n} \ea \ee
The iterated definitions (\ref{iter-n}) do not change if
we resp. apply $\hat\Delta,\Delta,\F\Delta $ to different tensor factors
[coassociativity of $\hat\Delta$; this follows from the coassociativity of
$\Delta$ and the cocycle condition (\ref{cocycle})];
for instance, for $n\!=\!3$ this amounts to (\ref{cocycle}) and
$\hat\Delta^{(3)}\!=\!(\hat\Delta\ot\id)\!\circ\!\hat\Delta$. For
any $g\in H[[h]]= \hH$ we shall use the Sweedler notations
$$
\ba{l}
\Delta^{(n)}(g)=\sum_I  g^I_{1} \otimes g^I_{2} \otimes ...
\otimes g^I_{n},\qquad\qquad
\hat\Delta^{(n)}(g)=\sum_I  g^I_{\hat 1} \otimes g^I_{\hat 2} \otimes ...
\otimes g^I_{\hat n}.
\ea
$$

\medskip 
\noindent 
Deforming the Euclidean  group $U\g^e$
by the twist (\ref{Moyalstarprod})$_2$ one finds $\beta\!=\!\1$, $\hat S\!=\! S$,
\bea 
&&\hat\Delta
(P_a)= P_a\ot\1+\1\ot P_a=\Delta (P_a),\nn &&\hat\Delta
(M_\omega)=M_\omega\ot\1+\1\ot M_\omega+
([\omega,\theta])^{ab}P_a\ot P_b
\nonumber 
\eea 
where $M_\omega\!=\!\omega^{ab}M_{ab}$ and $M_{ab}$ are the generators of $so(m)$; the Hopf subalgebra of translations is not deformed. 
Similarly one deforms  Poincar\'e transformations\cite{ChaKulNisTur04,Wes04,KocTso04}.

\medskip
\noindent
{\bf 2.2 \ Twisting $H$-module $*$-algebras.} \
A  left $H$-module $*$-algebra $\A$ is a
$*$-algebra equipped with a left action, i.e. a $\mathbb{C}$-bilinear map $(g,a)\!\in\!
H\!\times\!\A\!\to\! g\!\trc\! a\!\in\!\A$ such that
\be
(gg')\trc\! a=g \trc\! (g'\! \trc\! a)\!,\qquad (g\trc\!
a)^*\!=S(g)^*\trc\! a^*\!\!, \qquad g\trc\! (ab)=\!\sum_I
g^I_{1}\!\trc\! a\: \: g^I_{2}\!\trc\! b.        \label{leibniz}
\ee 
Given such an $\A$, let
$V\!\big(\A\big)$ the vector space underlying $\A$. \
 $V\!\big(\A\big)[[\lambda]]$ becomes a $\hH$-module $*$-algebra
$\A_\star$ when endowed with the product and $*$-structure
\be
 a\star a':= \left(\bFu  \trc a\right)
\left(\bFd  \trc a'\right),\qquad\qquad  
a^{*_\star}:=S(\beta)\trc a^*. \label{starprod} 
\ee
In fact, $\star$ is associative by (\ref{cocycle}),  fulfills $(a\!\star\!
a')^{*_\star}\!=\!a'{}^{*_\star}\!\!\star\! a^{*_\star}$ and
\be 
(g\trc\!
a)^{*_\star}\!=\hat S(g)^*\trc\! a^{*_\star},\qquad\qquad  
g\trc (a\!\star\!a')=\sum_Ig^I_{\hat
1}\!\trc\!  a\:\star\: g^I _{\hat 2}\!\trc\! a'. 
\ee
For the Moyal twist (\ref{Moyalstarprod}) $\beta\!=\!\1$, $*_\star\!=\!*$.
The {\it $\star$ is ineffective if $a$ or $a'$ is $H\!_s$-invariant:}
\be
g\trc a=\epsilon(g) a\quad \mbox{or}\quad g\trc a'=\epsilon(g)
a'\quad \forall g\in H\!_s \qquad\quad \Rightarrow\qquad \quad a\star
a'=aa'.\qquad \label{Trivstar}
\ee
Given $H$-module $*$-algebras $\A,\B$, also $\A\ot\B$ is,
so (\ref{starprod}) makes $V(\!\A\ot\B)$
into a $\hH$-module $*$-algebra $(\A\ot\B)_\star$.  Denoting
$a\ot\!_\star b\!=\!(a\ot\1_{\scriptscriptstyle \B})\!\star\!
(\1_{\scriptscriptstyle {\cal A}}\ot b)$, $\R\!=\!\Ru\!\ot\Rd$,  one finds
\be 
\ba{l} (a\ot_\star b)\star (a'\ot_\star
b')= a\star (\Rd\trc a') \ot_\star (\Ru\trc b)
\star b', \label{braid} \ea
\ee 
so $\ot_\star$ is the {\it braided tensor product} associated
 to $\R$, and $(\A\ot\B)_\star=\A_\star\ot_\star\B_\star$.

\noindent
{\bf If $\A$ is defined by generators $a_i$ and relations}, 
{\bf then so is $\A_\star$}, and fulfills PBW\cite{Fio10}. The {\it generalized Weyl map} 
is the linear map  
\ $\wedge:f\!\in\!\A\!\to\!\hat f\!\in\!\A_\star$  defined by 
\be 
f(a_1,a_2,...)\star=\hat f(a_1\star,a_2\star,...) \qquad\quad\mbox{in }
V(\A)=V(\A_\star),\label{defhat} 
\ee 
generalizing (\ref{defhatx}). It  fulfills 
$\wedge(ff')\!=\!\wedge(\Fu\trc\! f)\!\star\!\wedge(\Fd\trc\! f')$. 

\medskip
\noindent
{\bf 2.3 \ Deformation of the Heisenberg, Clifford algebra $\A^\pm$.}
Quantum mechanics on $\mathbb{R}^m$ is covariant w.r.t. the Lie group $G^e$
of  Euclidean - or, more generally, Galilei - transformations (here we 
consider them as active transformations). This
implies that the Heisenberg algebra $\A^+$ (resp. Clifford algebra 
$\A^-$) associated to a species of bosons (resp. fermions) is a $U\g^e$-module
$*$-algebra.  As the $G^e$-action is unitarily implemented on
the Hilbert spaces of the systems, that of $H=U\g^e$ is
defined on dense subspaces, in particular on a pre-Hilbert space
$\H$ of the one-particle sector, on which it will be denoted as
$\rho$: \
 $g \trc=: \rho(g)\!\in\!\O\!:=\!\mbox{End}(\H)$.

The pre-Hilbert space of $n$ bosons (resp. fermions) is described by
the completely symmetrized (resp. antisymmetrized) tensor product
$\H^{\ot n}_+$ (resp. $\H^{\ot n}_-$), which is a $H$-$*$-submodule
of $\H^{\ot n}$.  Assuming a unique, invariant vacuum state $\Psi_0$, the
bosonic (resp. fermionic) Fock space is defined as the closure
$\bH^{\infty}_{\pm}$ of
$$
\H^{\infty}_{\pm}:=\left\{\mbox{finite sequences
}(s_0,s_1,s_2,...) \in\mathbb{C}\Psi_0\oplus\H\oplus
\H^{2}_{\pm}\oplus...\right\}
$$
(finite
means that there exists an integer $l\!\ge\! 0$ such
that $s_n=0$ for all $n\!\ge\! l$). The
creation, annihilation operators $a^+_i,a^i$  associated to an orthonormal
basis $\{ e_i\}_{i\in\mathbb{N}}$ of $\H$ fulfill
the Canonical (anti)Commutation Relations (CCR) 
\be
a^ia^j=\pm a^ja^i,\qquad\quad
a^+_ia^+_j=\pm a^+_ja^+_i,\qquad\quad a^ia^+_j\mp a^+_ja^i=\delta^i_j\1_\A.
\label{ccr} 
\ee
($+$ for bosons, $-$ for fermions).\ $a^+_i,a^i$ resp. transform as
$e_i\!=\!a^+_i \Psi_0$ and $\la e_i,\cdot\ra$: 
\be 
g\trc e_i=\rho_i^j(g)e_j,\qquad 
g\trc a^+_i=\rho_i^j(g) a^+_j,\qquad g\trc a^i=\rho^\vee{}_i^j(g)a^j 
= \rho^i_j\big[S(g)\big] a^j      \label{lineartransf}
\ee 
($\rho^\vee\!=\!\rho^T\!\circ\! S$ is the contragredient
 of $\rho$). \ $\A^{\pm}$ are $H$-module $*$-algebras because the $\g^e$-action
(extended to products as a derivation) is compatible with (\ref{ccr}).

\medskip
Applying the deformation procedure one obtains $\hH$-module
$*_\star$-algebras $\A^{\pm}_\star$. The generators \ $a^+_i$, \ 
$a^{\prime i}\!:=\!a^+_i{}^{*_\star}\!=\!\rho^i_j(\beta)a^j$  \  fulfill
the $\star$-commutation relations  
\be\ba{l}
 a^{\prime i} \!\star\! a^{\prime j} =
\pm R^{ij}_{vu} a^{\prime u}\!\star\!  a^{\prime v} ,\\[8pt]
 a^+_i \!\star\! a^+_j = \pm R_{ij}^{vu} a^+_u \!\star\! a^+_v,\\[8pt]
 a^{\prime i}\!\star\!  a^+_j     = \delta^i_j\1_{\scriptscriptstyle{\cal A}}
\pm R^{ui}_{jv} a^+_u \!\star\! a^{\prime v}, \ea \quad
\Leftrightarrow\qquad  \ba{l} \hat a^{\prime i}\hat a^{\prime j} =
\pm R^{ij}_{vu}\hat a^{\prime u}\hat
a^{\prime v} ,\\[8pt]
\hat a^+_i\hat a^+_j = \pm R_{ij}^{vu}\hat a^+_u\hat a^+_v,\\[8pt]
\hat a^{\prime i}\hat a^+_j     = \delta^i_j\1_{\scriptscriptstyle
\hat{\cal A}} \pm R^{ui}_{jv}\hat a^+_u\hat a^{\prime v}, \ea
\label{hqccr} 
\ee 
where $R\!:=\!(\rho\ot \rho)(\R)$. The $a^{\prime i}$ transform 
according to the rule of the {\it twisted} contragredient representaton:  
$g\trc a^{\prime i}\!=\! \rho^i_j\big[\hat S(g)\big] a^{\prime j}$. Equivalently,
$\hA^\pm\sim\A^\pm_\star$ has generators $\hat a^+_i,\hat a^i$
fulfilling $\hat a^+_i{}^{\hat *}=\hat a^{\prime i}$ and the
rhs(\ref{hqccr})\cite{Fio96}. 

Is there a Fock-type representations of $\hA^\pm$? Yes, only one, on the 
{\it undeformed} Fock space of
bosons/fermions\cite{Fio10}. The important consequence is that {\bf
(\ref{hqccr}) are compatible with Bose/Fermi statistics}\cite{FioSch96}.
In fact one can {\it realize}\cite{Fio96} $\hat
a^+_i,\hat a^{\prime i}$ as  `dressed', hermitean conjugate  elements $\check a^+_i,\check
a^{\prime i}$ in $\A^\pm[[\lambda]]$ fulfilling (\ref{hqccr}): 
\be
\ba{l} \check a^+_i= \big(\bFu  \!\trc
a^+_i\big)\,\sigma\big(\bFd  \big),\qquad\qquad \check
a^{\prime i}= \big(\bFu  \!\trc a^{\prime
i}\big)\,\sigma\big(\bFd  \big). \label{defDf} \ea
\ee  
$\sigma$ is the generalized Jordan-Schwinger
realization of $U\g$, i.e. the $*$-algebra map
\ $\sigma\!:\!H[[\lambda]]\!\rightarrow\! \A^\pm[[\lambda]]$ \ such that 
\ $\sigma(\!\1_{\scriptscriptstyle
H}\!)\!=\!\1_{\scriptscriptstyle{\cal A}}$, $\sigma(g)\!=\!(g\trc
a^+_j) a^j$ if  $g\in \g$; it fulfills
$$
\ba{l}
g\trc a=\sum_I \sigma\big(g_{1}^I\big)\, a\,\sigma\left[S \big(g_{2}^I \big)\right]
\qquad\qquad\forall g\in H,\quad a\in \A^\pm.
\ea
$$
For $G=G^e$, and the Moyal twist
let $a^+_{\rm p},a^{\rm p}$ be the
creation, annihilation operators associated to the joint, generalized eigenvectors 
of the $P_a$, $P_ae_{\rm p}\!=\!p_ae_{\rm p}$ (${\rm p}\in\mathbb{R}^m$); in that basis 
$R^{{\rm p}{\rm p}'}_{qq'}=e^{i {\rm p}'\theta {\rm p}}\delta({\rm p}\!-\!{\rm q})\delta({\rm p}'\!-\!{\rm q}')$, where we abbreviate ${\rm p}\theta {\rm q}\!:=\!{\rm p}_a\theta^{ab} {\rm p}_b$, so that e.g. (\ref{hqccr})$_3$ becomes
 $\hat a^{{\rm p}}\hat a^+_{\rm q}     =\delta({\rm p}\!-\!{\rm q})\1_{\scriptscriptstyle
\hat{\cal A}} \pm e^{i {\rm p}\theta {\rm q}} \hat a^+_{\rm q}\hat a^{{\rm p}}$, and
(\ref{defDf}) becomes
$$
\ba{l} \check a^+_{\rm p}=a^+_{\rm p}e^{-\frac i2 p\theta\sigma(P)},
\qquad \qquad \check a^{\rm p}=a^{\rm p}e^{\frac i2
p\theta\sigma(P)},\qquad\qquad \sigma(P_a):=\int \!\!d^m\! p\: p_a
a^+_{\rm p} a^{\rm p}.
\ea
$$

\medskip
\noindent
{\bf 2.4 Moyal-deforming functions, differential, integral calculi on $\mathbb{R}\!^m$.}
We denote as  $\D_{p}$  
the Heisenberg algebra on $X=\mathbb{R}^m$.
The $*$-structure and the $\star$-commutation relations of  the $\hH$-module $*$-algebra  $\D^{\ot n}_{p\star}$ are as the undeformed ones except 
(\ref{summary}), where we have abbreviated $\rx^{h}_1$, $\rx^{h}_2$,...  for 
$\rx^{h}\ot \1\ot...$, $\1\ot \rx^{h}\ot...$ ,...  and $\partial_{\rx_1^a}=\partial/\partial \rx_1^a$,
$\partial_{\rx_2^a}=\partial/\partial \rx_2^a$,... \
$\1,\rx^{h}_1$, $\rx^{h}_2$,... generate the $\hH$-module $*$-subalgebra   
$\X^{\ot n}_{p\star}$. The latter is  'too small' for physical purposes, but one can extend $\star$ and the
Weyl map $\wedge$ to other $H$-module $*$-algebras, e.g. the Schwarz space $\X\!:=\!{\cal S}(\mathbb{R}^m)$, the distribution space $\X'$  and the algebra of $\D\supset\D_p$ of smooth differential operators on $X$, replacing the (discrete) set of polynomials in $\rx_i^a$ by the 
(continuous)  set of
exponentials $e^{i\sum_ih_i\cdot \rx_i}$ (labelled by $n$ indices
$h_i\!\in\!\mathbb{R}^m$) as a basis: since  the $\theta$-power series expansion for the
$\star$-product of two exponentials converges, giving in particular
 \be
e^{ih\cdot \rx_i}\star e^{ik\cdot \rx_j}
=e^{i\left(h\cdot \rx_i+k\cdot \rx_j-\frac{h\theta k}2\right)}
,   \label{expstarprod}
\ee
 it suffices to express the $a,b\!\in\!\X,\X'$ through their Fourier transforms $\tilde a, \tilde b$ and {\it define}
\be
a(\rx_i)\star b(\rx_j):= \int\! d^m\!h\!\int\!  d^m\!k\:\, e^{i\left(h\cdot \rx_i+k\cdot \rx_j-
\frac{h\theta k}2\right)}\tilde a(h) \tilde b(k).                        \label{IntForm}
\ee
The Moyal $\star$-product  fulfills the cyclic property w.r.t. Riemann integration:
\be
\int\!\!
d\rx  \, a(\rx)\star b(\rx)=\int\!\! d\rx \, a(\rx)\,b(\rx)=
\int\!\! d\rx \, b(\rx) \star a(\rx)            \label{intprop}
\ee
(this is modified\cite{Fio10} when $\beta\neq\1$); the same applies for multiple integrations, 
after having reordered through (\ref{braid}) all the functions depending on the same argument 
beside each other, e.g. 
$f(\rx_i)\star  f'(\rx_j)\star  f''(\rx_i)=  
\Ru\!\trc\! [f'(\rx_j)]\star \Rd\!\trc\! [f(\rx_i)]\star f''(\rx_i)$.
One can define also a $\hH$-invariant ``integration over $\hat
X$'' $\int\! d\hrx$ such that for each
$f\!\in\!\X$ 
\be 
\int\!\! d\hrx\, \hat
f(\hat {\rm x}) =\int\!\! d{\rm x}\, f({\rm x}).   \label{hatint}
\ee
We shall call $\wedge^n$ the analogous maps 
$\wedge^n:f\!\in\!\X^{\ot n}[[\lambda]]\!\to\!\hat f\!\in\!(\X^{\ot n})_\star$. 
One finds 
$\wedge(e^{ih\cdot \rx_i})\!=\!e^{ih\cdot \rx_i\star}\!\leadsto\! e^{ih\cdot \hrx_i}$.
Eq. (\ref{intprop}-\ref{hatint}) generalize to integration over $n$
independent variables.
The differences \ $\xi^a_i\!:=\! \rx^a_i\!-\! \rx^a_{i\!+\!1}$, \
$i\!=\!1,...,n\!-\!1$, \ are translation invariant, so by (\ref{Trivstar}) 
$f(\rx)\star h(\xi)=f(\rx)h(\xi)=h(\xi)\star f(\rx)$ for all $f,h$.

 \section{Twisting non-relativistic second quantization}\label{se:2}

In the wave-mechanical description of a system of $n$ bosons/fermions
 on  $\mathbb{R}^m$ we describe any abstract state vector (ket)
$s\!\in\!\H^{\ot n}_\pm$ as
a smooth wavefunction $\psi_s\!\in\!\X^{\ot n}_\pm$ of $\rx_1,...,\rx_n$.
By the Weyl map we can describe the same state $s$ also as the noncommutative 
wavefunction $\wedge(\psi_s)\!\equiv\!\hat\psi_s\!\in\!\widehat{\X^{\ot n}_{\pm}}$
of $\hrx_1,...,\hrx_n$.
The maps $s\stackrel{\kappa^n_\pm}{\longrightarrow}\psi_s \stackrel{\wedge^n}{\longrightarrow}\hat\psi_s$ are unitary.
Differential operators $D\!\in\!\D^{\ot n}_+$ acting on $\X^{\ot n}_\pm$ are
mapped into differential operators 
$\hat D\!\in\!\widehat{\D^{\ot n}_+}$ acting on $\widehat{\X^{\ot n}_{\pm}}$.
The action of the symmetric group $S_n$ on $\widehat{\X^{\otimes n}}$
is obtained by ``pull-back'' from that on $\X^{\otimes n}$: a
permutation $\tau\!\in\!S_n$ is represented on $\X^{\otimes
n}\!,\!\widehat{\X^{\otimes n}}$ resp. by the permutation operator
$\P_\tau$ and the ``twisted permutation operator''
$\P^F_\tau\!=\!\wedge^n \P_\tau [\wedge^n]^{-1}$.
Thus, $\widehat{\X^{\otimes n}_{\pm}}$ are
(anti)symmetric  up to the similarity transformation $\wedge^n$ (cf.
Ref. \cite{FioSch96}).
$H$-equivariance of the commutative description translates into 
$\hH$-equivariance of the noncommutative one. 
Given a basis $\{ e_i\}_{i\!\in\!\mathbb{N}}\!\subset\!\H$, let $\varphi_i\!=\!\kappa(e_i)$,
 $\hat\varphi_i\!=\!\wedge(\varphi_i)$;
we illustrate how $\wedge^2$
transforms the (anti)symmetrized tensor product basis: 
\be
\varphi_{i}(\rx_1\!) \varphi_{j}(\rx_2\!)\!\pm\!\varphi_{j}(\rx_1\!) \varphi_{i}(\rx_2\!)
\stackrel{\wedge^2}{\longrightarrow}F^{hk}_{ij}
\hat\varphi_{h}(\hat{\rm x}_1) \hat\varphi_{k}(\hat{\rm x}_2)\!\pm\!
F^{hk}_{ji}\hat\varphi_h(\hat{\rm x}_1) \hat\varphi_{k}(\hat{\rm x}_2)
\ee
where $F\!:=\!(\rho\ot\rho) (\F)$. To make at most the matrix $R\!:=\!(\rho\ot\rho) (\R)$ 
 appear at the rhs one should rather use at the lhs
as  vectors of a (non-orthonormal) basis of $(\X\ot\X)_\pm$
\bea
\overline{F}^{hk}_{ij}[\varphi_h(\rx_1\!) \varphi_k(\rx_2\!)\!\pm\!\varphi_k(\rx_1\!) 
\varphi_h(\rx_2\!)] &\stackrel{\wedge^2}{\longrightarrow}&
\hat\varphi_{i}(\hat{\rm x}_1) \hat\varphi_{j}(\hat{\rm x}_2)\!\pm\!
R^{{k}{h}}_{{i}{j}}\hat\varphi_h(\hat{\rm x}_1) \hat\varphi_{k}(\hat{\rm x}_2) \label{n2}\\
&&=\hat\varphi_{i}(\hat{\rm x}_1) \hat\varphi_{j}(\hat{\rm
x}_2)\!\pm\!\hat\varphi_{i}(\hat{\rm x}_2) \hat\varphi_{j}(\hat{\rm x}_1). \label{n2'}
\eea
Their form (\ref{n2'}) is closer than (\ref{n2}) to the undeformed
counterpart. Both generalize  to $n>2$. The
generalization of (\ref{n2'}) for $n$  fermions is the Slater determinant
\be
\psi^{(n)}_{-,i_1...i_n}(\hrx_1,...,\hrx_2)=...\left\vert\ba{llll}
\hat\varphi_{i_1}(\hat{\rm x}_1) \:&\: \hat\varphi_{i_2}(\hat{\rm x}_1) \:&\: ... \:&\: \hat\varphi_{i_n}(\hat{\rm x}_1) \\
\hat\varphi_{i_1}(\hat{\rm x}_2) \:&\: \hat\varphi_{i_2}(\hat{\rm x}_2) \:&\: ... \:&\: \hat\varphi_{i_n}(\hat{\rm x}_2)\\
... \:&\: ... \:&\: ... \:&\: ...\\
\hat\varphi_{i_1}(\hat{\rm x}_n) \:&\: \hat\varphi_{i_2}(\hat{\rm x}_n) \:&\: ... \:&\: \hat\varphi_{i_n}(\hat{\rm x}_n)
\ea\right\vert ,
\ee
provided we keep the order of the wavefunctions and permute the $\hrx_h$:
to the permutation $(h_1,\!h_2,\!...,\!h_n)$ there corresponds the term
$\pm\,\hat\varphi_{i_1}(\hat{\rm x}_{h_1})\hat\varphi_{i_2}(\hat{\rm x}_{h_2}) ... \hat\varphi_{i_n}(\hat{\rm x}_{h_n})$.

\noindent
The {\bf nonrelativistic field operator}  and its hermitean
conjugate (in the Schr\"odinger picture) 
\be 
\varphi({\rm x}):= \varphi_i({\rm x})a^i,
\qquad\qquad\qquad\qquad \varphi^*({\rm x})\!=\! \varphi_i^*({\rm x}) a^+_i 
\ee 
(infinite sum over $i$) are operator-valued
distributions fulfilling the CCR
\be 
[\varphi({\rm
x}),\varphi({\rm y})]_{{\scriptscriptstyle \mp}}=\mbox{h.c.}=0,
\qquad \quad [\varphi({\rm x}),\varphi^*({\rm
y})]_{{\scriptscriptstyle \mp}}= \varphi_i({\rm x})\varphi_i^*({\rm
y})=\delta({\rm x}\!-\!{\rm y}) \label{fccr} 
\ee 
($\mp$ for bosons/fermions). The {\it field $*$-algebra} $\Phi$ 
is spanned by all monomials 
\be 
\varphi^*({\rm x}_1)....\varphi^*({\rm x}_m)
\varphi({\rm x}_{m\!+\!1})...\varphi({\rm x}_n) \label{fieldmon}
\ee
(${\rm x}_1,...,{\rm x}_n$ are independent points). So $\Phi\subset
\Phi^e\!:=\!\A^\pm\ot\left(\bigotimes_{i=1}^{\infty}\!
\X'\right)$. Here the 1st, 2nd,... tensor factor
$\X'$ is the space of distributions depending on ${\rm x}_1,{\rm
x}_2,\!...$; the dependence of 
(\ref{fieldmon}) on ${\rm x}_{h}$ is trivial for $h> n$. $\Phi^e$ is a huge $H$-module $*$-algebra:
$a^+_i,\varphi_i$ transform as $e_i$, and $a^i,\varphi_i^*$
transform as $\la e_i,\cdot\ra$.
The CCR (\ref{ccr}) of $\A^\pm$ are the only nontrivial commutation 
relations in $\Phi^e$.
A key property is that $\varphi,\varphi^*$ are basis-independent,
i.e. {\bf invariant under the group $U(\infty)$ of unitary
transformations of $\{ e_i\}_{i\!\in\!\mathbb{N}}$}, in particular under
the subgroup $G^e$ of Euclidean transformations (transformations
of the states $e_i$ obtained by translations or rotations of the
1-particle system), or (in infinitesimal form) 
{\bf under $U\g^e$}: $g\trc \varphi({\rm
x})=\epsilon(g)\varphi({\rm x})$.
By (\ref{Trivstar}), deforming $U\g^e\to\widehat{U\g^e}$, $Uu(\infty)\to\widehat{Uu(\infty)}$
and $\Phi^e\stackrel{(\ref{starprod})}{\longrightarrow}\Phi^e_\star\sim\widehat{\Phi^e}$,
we still find for all $\omega\!\in\!V(\Phi^e)[[\lambda]]$
\be
\varphi({\rm x})\!\star\! \omega=\varphi({\rm x}) \omega, \qquad
\omega\!\star\!\varphi({\rm x})=\omega\,\varphi({\rm x}), \qquad \quad
\&\quad  \mbox{h. c.}.             \label{fieldinv} 
\ee
Since $\epsilon(\beta)=1$ and the definition
$a^{\prime i}\!:=\!a^+_i{}^{*_\star}\!=\!S(\beta)\trc a^i$ imply
\be
\varphi({\rm x}) =\varphi_i({\rm x}) \star a^{\prime i},
\qquad\qquad\varphi^*({\rm x})=\varphi^{*_\star}({\rm x}) =a^+_i\star \varphi_i^{*_\star}({\rm x}), \ee
and $\varphi_i({\rm x})\varphi_i^*({\rm y})=\varphi_i({\rm
x})\star\varphi_i^{*_\star}({\rm y})$, in $\Phi^e_\star$  the CCR (\ref{fccr}) become 
\be 
[\varphi({\rm x})\stackrel{\star}{,}\varphi({\rm
y})]_{\mp}=h.c.=0, \qquad \qquad\qquad [\varphi({\rm
x})\stackrel{\star}{,}\varphi^{*_\star}({\rm y})]_{\mp}=\varphi_i({\rm x})\star\varphi_i^{*_\star}({\rm y})\qquad\quad
 \label{qfccr}
\ee (here $[A\!\stackrel{\star}{,}\!B]_{\mp}\!:=\!A\star
B\!\mp\! B\star A$). $\Phi^e_\star$ is a huge
$\widehat{U\g^e}$- [and $\widehat{Uu(\infty)}$-] module
$*$-algebra.

The unitary
map $\hat \kappa^n_{\pm}=\wedge\circ\kappa^n_{\pm} :s\!\in\!\H^{\ot n}_{\pm}\leftrightarrow
\hat \psi_s\!\in\!\widehat{\X^{\ot n}_{\pm}}$ and its inverse can be expressed using
the field, as in the undeformed theory\cite{Fio10}. For instance,
$\hat \kappa(e_i)=\hat\varphi_i(\hat{\rm x})=
\left\la\Psi_0,\hat\varphi(\hat{\rm x})\hat a^+_i \Psi_0\right\ra$
and the wavefunction (\ref{n2'}) equals
$\left\la\Psi_0,\hat\varphi(\hat{\rm
x}_2)\hat\varphi(\hat{\rm x}_1)\hat a^+_{i}\hat a^+_{j}\Psi_0\right\ra$.

\noindent
Assume the $n$-particle wavefunction $\psin$ fulfills the
 Schr\"odinger eq. (\ref{1Schr}) if $n\!=\!1$, and 
\be \ba{l}
i\hbar \frac{\partial}{\partial t}\psin=\Hans\psin,\qquad \quad
\Hans\!:= 
\sum\limits_{h=1}^{n}\haus({\rm x}_h,\partial_h,t)\star +\sum\limits_{h<k}
W(|\rx_h\!-\!\rx_k|)    
\star\ea \label{Schr}
 \ee
if $n\!\ge\! 2$; here we keep the time coordinate $t$  ``commuting'. $\Hans$ is
hermitean if $\hau$ is and $\beta\!\trc\! \hau\!=\!\hau$, 
as we shall assume. In general, (\ref{Schr}) is a
$\star$-differential, pseudodifferential equation preserving the
(anti)symmetry of $\psin$. The Fock space Hamiltonian
\be
\Has= \int\!\!\!\! d{\rm x} \varphi^{*_\star}({\rm
x})\star\hau\star \varphi({\rm x})\!\star  + \!\int \!\!
\!\!d{\rm x}\!\!\! \int \!\!\!\! d{\rm y}
\varphi^{*_\star}({\rm y})\!\star\!\varphi^{*_\star}({\rm x})\!\star\!W(|\hrx\!-\!\hat{\rm y}|)\!\star\!
\varphi({\rm x})\!\star\!\varphi({\rm y})\star
\ee
annihilates the vacuum, commutes with the number-of-particles
operator $\mbox{\bf n}\!:=\!a^+_i\!\star a^i$ and its restriction to
$\H^{\ot n}_{\pm}$ coincides with $\Hans$ up to the unitary
transformation $\tilde\kappa^{\ot n}$. As in the undeformed theory,
 formulating the dynamics
on the Fock space allows to consider also more general Hamiltonians
$\Has$, which do not commute with $\mbox{\bf n}$.
Replacing $\hat V({\rm x}\star,t)=V({\rm x},t)\star$, $\hat {\rm
A}({\rm x}\star,t)={\rm A}({\rm x},t)\star$, $\hat\varphi_i({\rm
x}\star)=\varphi_i({\rm x})\star$ we can reformulate the previous
equations  within $\hat\Phi^e$, $\hat\Phi$ using only $\star$-products,
or equivalently, dropping $\star$-symbols and using only
``hatted'' objects:
\be
\ba{l}\hat\varphi(\hat {\rm x}) =\hat\varphi_i(\hat {\rm x})
\hat a^{\prime i},\qquad \qquad\qquad\qquad\:\:\:\varphi^{\hat
*}(\hat {\rm
x}) =\hat a^+_i \hat \varphi_i^{\hat *}(\hat {\rm x}) \\[10pt]
[\hat \varphi(\hat {\rm x}),\hat \varphi(\hat {\rm
y})]_{\mp}=\mbox{h.c.}=0, \qquad \qquad\quad [\hat \varphi(\hat
{\rm x}),\hat \varphi^{\hat
*}(\hat {\rm y})]_{\mp}=\hat\varphi_i(\hat{\rm
x})\hat\varphi_i^{\hat *}(\hat{\rm y}),\\[10pt]
i\hbar \frac{\partial}{\partial t}\hat\psi^{(n)}=\hat\Ha^{(n)}\hat\psi^{(n)},
\qquad\qquad\qquad\hat \Ha^{(n)}\!=\! 
\sum\limits_{h=1}^{n}\!\hat\hau
(\hat{\rm
x}_h,\hat\partial_h,t)\!+\!\sum\limits_{h<k}\! \hat W(|\hrx_h\!-\!\hrx_k|),\\[12pt]
\hat\Ha=\!\displaystyle\int \!\!\! d\hrx
\hat \varphi^{\hat *}(\hat {\rm x})\hhau^{}\!(\hat{\rm
x},t)\hat\varphi(\hat{\rm x})  + \!\!\int \!\! \! d\hrx\!\! 
\int \!\!\! d\hat {\rm y}
\hat\varphi^{\hat *}\!(\hat {\rm y})\hat\varphi^{\hat *}\!(\hat {\rm x})
W(|\hrx\!-\!\hat{\rm y}|) \hat\varphi({\rm x})\hat\varphi({\rm y}).
\ea\label{hatqfccr} 
\ee
As in the undeformed
case, the field in the Heisenberg picture fulfills\cite{Fio10} the equation of motion 
$i\hbar \frac{\partial}{\partial t}\hat\varphi_{\scriptscriptstyle H}= [\hat\varphi^{\scriptscriptstyle H},\hat\Ha]$ and (at equal times) the (anti)commutation
relations (\ref{hatqfccr})$_{3,4}$, where the rhs is a ``$c$-number'' distribution.
Formulae  (\ref{hatqfccr}), with the related ones
for the Heisenberg field $\hat\varphi_{\scriptscriptstyle H}$, summarize
our framework for a {\it $\hH$-covariant nonrelativistic 
field quantization on the noncommutative spacetime
$\mathbb{R}^m_\theta$ compatible with the axioms of quantum mechanics,
including Bose and Fermi statistics}. 

\medskip
If $\hau,W$ are $H_s$-invariant then
$\Hans=\Han,\:\Ha_s=\Ha$, i.e. the dynamics is not deformed, and
the total energy is additive if $W\equiv 0$. 
As we now show,  if  $W\equiv 0$
but $\hau$ is {\it not} $H\!_s$-invariant then additivity fails,
i.e. \ $\Hans\neq \sum_{h=1}^{n}\haus({\rm x}_h,\partial_h,t)$: \
so to say, a mutual interaction among the particles is built in due to 
the $\star$-products. 
We just consider a system of 2 bosons/fermions in  a perpendicular magnetic field $B=$const
on the 2-dimensional Moyal space; this means that
$\theta^{ab}\!=\!\theta\epsilon^{ab}$ and that in the Hamiltonian $\haus$
of (\ref{1Schr})  \ $V\!\equiv\! 0$ and, choosing the  symmetric gauge for the vector
potential,
$A^a({\rm x})\!=\!-B\epsilon^{ab}{\rm x}^b/2$, implying 
$D_a=\partial_a\!-\!ib\,\epsilon^{ab}{\rm x}^b$, $b\!:=\!\frac{qB}{2\hbar c}$.  One
finds 
\be
\haus(\rx,\partial_\rx)=-\frac{\hbar^2}{2m}D_a \star D_a\star
=\frac{\hbar^2}{2m}\!\left[-\!\left(1\!+\!\frac{b\theta}2\right)^2\!\!
\Delta+b^2  {\rm x}^2-2b\!\left(1\!+\!\frac{b\theta}2\right)  l \right]
\ee
where $l=-i\epsilon^{ab}\rx^a\partial_b= i({\rm x}^2\partial_1\!-\!{\rm x}^1\partial_2)$ is the 
angular momentum in dimensionless units. So $\haus$ can be obtained from $\hau$ rescaling 
the coordinates by $\sqrt{1\!+\!\frac{b\theta}2}$ and multiplying the result
by  $1\!+\!\frac{b\theta}2$. \ $\haus$ has Landau-type 
levels and eigenfunctions, more easily expressible\cite{Fio10} using the $\hrx$. 
If $W\!=\!0$ we find instead by an easy computation
\be
\Has^{(2)}
=\haus(\rx_1,\partial_{x_1})\!+\!\haus(\rx_2,\partial_{x_2})\!+\!
\frac{\hbar^2b\theta}{2m}\!\left[ib\epsilon^{ab}(\rx^a_1\partial_{\rx_2^b}
\!+\!\rx^a_2\partial_{\rx_1^b})\!-\!(2\!+\!b\theta)  \partial_{\rx_1^a}
\partial_{\rx_2^a} \right].
\ee
The last term breaks the additivity of $\Has^{(2)}$, as claimed.


\end{document}